\documentclass[pre]{revtex4}
\usepackage{amsmath}
\usepackage{graphicx}
\begin{document}
\title{Slip-Size Distribution and Self-Organized Criticality in Block-Spring Models with Quenched Randomness}
\author{Hidetsugu Sakaguchi and Shuntaro Kadowaki}
\affiliation{Department of Applied Science for Electronics and Materials,
Interdisciplinary Graduate School of Engineering Sciences, Kyushu
University, Kasuga, Fukuoka 816-8580, Japan}
\begin{abstract}
We study  slowly pulling block-spring models in random media. Second-order phase transitions exist in a model pulled by a constant force in the case of velocity-strengthening friction. If external forces are slowly increased, nearly critical states are self-organized. Slips of various sizes  occur, and the probability distributions of slip size roughly obey power laws. The exponent is close to that in the quenched Edwards--Wilkinson model. Furthermore, the slip-size distributions are investigated in cases of Coulomb friction and velocity-weakening friction.
\end{abstract}
\maketitle
\section{Introduction}
Power laws are observed in various research fields. 
One of the most well-known power laws is the Gutenberg--Richter law, which states  that  the energy released during earthquakes obeys a power law.~\cite{rf:1,rf:2} That is, the probability distribution of the slip size of faults obeys a power law of exponent around $\tau=1.67$.  
Burridge and Knopoff proposed a block-spring model of earthquakes.~\cite{rf:3} Later, Carlson and Langer used a type of velocity-weakening friction and observed a power law in the range of small magnitudes in a homogeneous block-spring model.~\cite{rf:4} Many authors have studied extensively the Carlson-Langer model and its modification.~\cite{rf:5,rf:6,rf:7,rf:8} 
There are also some experimental studies on stick-slip motion and the power-law behaviors.~\cite{rf:9,rf:10}  

On the other hand, Bak et al. proposed a cellular automaton model for sand piles, which exhibits self-organized criticality (SOC).~\cite{rf:11} Bak et al. insisted that the SOC is related to the Gutenberg--Richter law for earthquakes. However, avalanches in real sand piles hardly exhibit a power law in most cases.  Frette et al. found that rice piles in a Hele-Shaw cell exhibit SOC and proposed a cellular automaton model for rice piles.~\cite{rf:12} In the Oslo rice pile model, a rice grain is placed on the boundary site, and  repeated toppling occurs if the height difference between neighboring sites is beyond a critical value. The critical value takes a random value depending on the site.  The probability distribution of the avalanche size takes a power law of exponent $\tau\sim 1.55$ in the rice pile model. 

There is a relationship between the rice pile model and the interface growth model. The interface growth problem has been another topic in the study of nonequilibrium phenomena. There are a number of applications of random interface-growth problems such as bacterial colony growth and fluid invasion into porous media~\cite{rf:13,rf:14}. In many experiments, the randomness is fixed in space.  One model for interface growth with such quenched disorder is the KPZQ equation~\cite{rf:15,rf:16}, which is a modification of the Kardar-Parisi-Zhang (KPZ) equation~\cite{rf:17}. The KPZQ model becomes the quenched Edwards--Wilkinson (qEW) equation~\cite{rf:18} if the nonlinear term is absent. The Oslo rice pile model can be mapped to a kind of qEW equation. Paczuski and Boettcher showed that the slip-size distribution of the qEW equation driven at the boundary exhibits a power law of exponent $\tau\sim 1.55$.~\cite{rf:19} The two models are considered to belong to the same universality class. 

In this paper, we study a block-spring model with quenched randomness and discuss the slip-size distribution. The behavior of block-spring models with quenched randomness is rather different from that of the homogeneous block-spring models. This model has a similarity to the random interface growth model, and we will discuss the relationship among the block-spring models with quenched randomness, random interface models, and SOC.  In Sect.~2, we explain the self-organized criticality in the random interface model in more detail. In Sect.~3, we show numerical results of the block-spring model with quenched randomness. In Sect.~4, we summarize the numerical results. 

\section{Random Interface Growth Model and Self-Organized Criticality}
The KPZQ equation in one dimension is written as 
\begin{equation}
\frac{\partial z}{\partial t}=k \frac{\partial^2z}{\partial x^2}+\lambda\left(\frac{\partial z}{\partial x}\right )^2+f+\eta(x,z),
\end{equation}
where $\eta(x,z)$ denotes the quenched disorder satisfying $\langle \eta\rangle=0$ and $\langle \eta(x,z)\eta(x^{\prime},z^{\prime})\rangle=2D\delta(x-x^{\prime})\delta(z-z^{\prime})$. 
If $\lambda=0$, the KPZQ equation becomes the qEW equation.

The KPZQ equation exhibits a pinning-depinning transition. When the driving force $f$ is larger than the critical value $f_c$, 
the interface grows with a certain average velocity. When $f$ is below $f_c$, the interface is pinned in the quenched random medium. In random interface problems, the  root-mean-square width $W(l)=\langle (z(x)-\langle z\rangle)^2\rangle^{1/2}$ for $0<x<l$ is often investigated.   
At the critical driving force $f_c$, a scaling law $W(l)\sim l^{\alpha}$ is satisfied, where the exponent $\alpha$ is evaluated at $0.63$. The exponent $\alpha$ is related to two exponents $\nu_{\|}$ and $\nu_{\bot}$ in the directed percolation problem as $\alpha=\nu_{\|}/\nu_{\bot}$, where $\nu_{\|}$ is the exponent for the longitudinal correlation length $\xi_{\|}$ and $\nu_{\bot}$ is the exponent for the transverse correlation length $\xi_{\bot}$.  The scaling law is theoretically satisfied only at the critical driving force $f_c$. The exponent $\alpha=0.63$ is close to the experimentally evaluated value in the paper-wetting experiment by Buldyrev et al.~\cite{rf:20}.  In their experiment, a paper is vertically immersed in a container of ink, and the interface of the dry and wet regions rises randomly. The scaling law was observed even if control parameters were not precisely adjusted to the critical value.  

We have proposed a model equation in which the critical condition is naturally obtained even when the initial driving force deviates from the critical value~\cite{rf:21}. The model equation is written as 
\begin{equation}
\frac{\partial z}{\partial t}=k \frac{\partial^2z}{\partial x^2}+\lambda\left(\frac{\partial z}{\partial x}\right )^2+F-d z+\eta(x,z),
\end{equation}
where the term $F-dz$ implies that the driving force decreases in proportion to $z$. If the term $dz$ is replaced with the spatial average $d\langle z\rangle$,  where $\langle z(t)\rangle$ is the spatial average of the interface height in system size $L$: $\langle z(t)\rangle=(1/L)\int_0^Lz(x,t)dx$,
Eq.~(2) becomes  
\begin{equation}
\frac{\partial z}{\partial t}=k \frac{\partial^2z}{\partial x^2}+\lambda\left(\frac{\partial z}{\partial x}\right )^2+F-d\langle z\rangle+\eta(x,z).
\end{equation}
This model equation is closely related to the KPZQ equation Eq.~(1), or it can be interpreted as a modified KPZQ equation with a feedback term. 
Considering the paper-wetting experiment, we interpret $z(x,t)$ as the height of the interface at the horizontal position $x$. 
If $F>f_c$, the interface rises from the flat initial condition $z(x,0)=0$. The driving force becomes weaker owing to the term of $F-d\langle z\rangle$ as the average height $\langle z\rangle$ increases. In the paper-wetting problem, the driving force becomes weaker because of gravity and the evaporation effect from both sides of the paper as the interface rises.    
The interface growth stops owing to the pinning transition when $F-d\langle z\rangle\simeq f_c$ is satisfied. 
After the interface is pinned, the critical condition $F-d\langle z\rangle\simeq f_c$ is maintained forever because  $\langle z\rangle$ is constant in time. Thus, a nearly critical state is self-organized by this very simple mechanism. We have performed a direct numerical simulation and confirmed that SOC is realized and  that $W(l)$ satisfies that $W(l)\sim l^{\alpha}$ with $\alpha\simeq 0.64$.  

We can further generalize the model equations in Eqs.~(2) and (3) as
\begin{equation}
\frac{\partial z}{\partial t}=k \frac{\partial^2z}{\partial x^2}+\lambda\left(\frac{\partial z}{\partial x}\right )^2+v_0t-dz+\eta(x,z)
\end{equation}
and
\begin{equation}
\frac{\partial z}{\partial t}=k \frac{\partial^2z}{\partial x^2}+\lambda\left(\frac{\partial z}{\partial x}\right )^2+v_0t-d\langle z\rangle+\eta(x,z),
\end{equation}
where $v_0$ is sufficiently small. The driving force increases slowly with the term $v_0t$. The total effective driving force $v_0t-d\langle z\rangle$ increases slowly beyond $f_c$ but decreases by the rising of the interface. The effective driving force $f=v_0t-d\langle z\rangle$ fluctuates around  $f_c$ by the intermittent rising motion or slip process. 
In the experiment of paper-wetting, this slow increase in the driving 
force can be realized if the paper is slowly pushed down under the water surface. The random interface is forced to move forward slowly on average, if the interface is measured from the bottom of the paper.  The interface remains almost stationary when the interface is pinned to the random medium; however, it occasionally exhibits large slips owing to the depinning from the random medium. As a result, the average velocity of the interface is $d\langle z\rangle/dt \sim v_0/d$. On the other hand, the average velocity increases continuously from 0 as $v\sim (f-f_c)^{\beta}$ for  $f>f_c$ in the original KPZQ equation Eq.~(1). If $v_0$ is set to be a sufficiently small value, $v_0t-d\langle z\rangle$ fluctuates around $f_c$ as $(f-f_c)^{\beta}=(v_0t-d\langle z\rangle -f_c)^{\beta}$ becomes the same order as $v_0/d$. Then, the probability distribution of the slip size is expected to obey a power law because a nearly critical state is realized in this slowly driven system. 

\begin{figure}[tbp]
\begin{center}
\includegraphics[height=3.5cm]{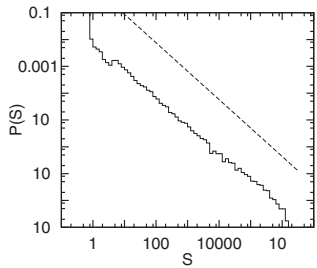}
\end{center}
\caption{Slip-size distribution $P(S)$ at $\lambda=0$. The dashed line denotes a power law of exponent $\tau=1.06$. }
\label{f1}
\end{figure}

To demonstrate the nearly self-organized critical state in a slowly driven system, we show numerical results of discretized models for Eq.~(5) expressed as
\begin{equation}
\frac{\partial z_i}{\partial t}=k (z_{i+1}-2z_i+z_{i-1})+(\lambda/4)(z_{i+1}-z_{i-1})^2+v_0t-d \langle z\rangle +\eta(i,z_i),
\end{equation}
where $\langle z\rangle=(1/N)\sum_{j=1}^Nz_j$, $k=5,\lambda=0$, $F=0.15,v_0=0.005$, $d=0.00005$, and $N=10000$. The quenched random force $\eta(i,z_i)$  is assumed to 
take a uniform random number between -3 and 3. Furthermore, we have used a stepwise time evolution. That is, the numerical simulation of Eq.~(6) is performed for a fixed value of $v_0t-d\langle z\rangle$ until $dz_i/dt$ becomes 0 for every $i$. After the relaxation process, $v_0t$ is increased to $v_0(t+\Delta t)$. At each step, $z_i$ moves to $z_i+\Delta z_i$ and the total slip size $S=\sum_{j=1}^N\Delta z_j$ is calculated.  The critical value $f_c$ for the discretized KPZQ equation Eq.~(1) is $f_c\simeq 0.792$ for $k=5$ and $\lambda=0$.  In numerical simulation of Eq.~(6), $v_0t-d\langle z\rangle$ fluctuates around $0.8$. This means that a nearly critical state is naturally self-organized. That is, the nearly critical state is an attractor in this slowly driven system.    Figure 1 shows the probability distribution of $S$, which is approximated by the power law $P(S)\propto 1/S^{\tau}$ with $\tau=1.06$. Several authors evaluated the exponents of the slip-size distribution as $\tau\sim 1.08$ for the qEW equation using constant-velocity ensembles.~\cite{rf:22,rf:23} The exponent shown in Figs.~1 is consistent with the exponents reported in these previous works. 

\section{Block-Spring Model with Quenched Randomness} 
The block-spring model for earthquakes is expressed as 
\begin{equation}
\frac{d^2z_i}{dt^2}=k(z_{i+1}-2z_i+z_{i-1})+v_0t-dz_i+\eta(v_i),
\end{equation}
where $k(z_{i+1}-2z_i+z_{i-1})$ denotes the interaction between neighboring blocks with spring constant $k$, $v_0t-dz_i=d(v_0t/d-z_i)$ denotes the coupling to a rigidly moving plate of constant velocity $v_0/d$ with spring constant $d$, and  $\eta(v)$ represents a kinetic friction depending on the velocity $v=dz/dt$. Carlson and Langer used a velocity-weakening friction: $\eta(v)=-1/(1+\gamma v)$. The velocity-weakening friction with $\gamma>0$ makes a uniform state unstable and spatio-temporal chaos appears as a result of the instability. In the numerical simulation by Carlson and Langer, a power law of $\tau\sim 1$ was observed for small slip size. However, large-scale slips occur rather frequently and a hump structure appears in the slip-size distribution for $\gamma>4$.~\cite{rf:4} We have performed numerical simulation of the block-spring models on a percolation cluster and found a power law of $\tau\sim 1.67$ when the parameter is around a specific value of $\gamma=2$.~\cite{rf:8} For $\gamma>2$, a hump structure is seen even in our numerical simulation on a percolation cluster, and  the slip-size distribution decays rapidly above a certain size and does not obey a power law for $\gamma<2$.  

We can consider another block-spring model equation in which the external pulling force is constant and quenched randomness is involved as  
\begin{equation}
\frac{d^2z_i}{dt^2}=k(z_{i+1}-2z_i+z_{i-1})+f-\nu_0-\nu_1v_i+\eta(i,z_i),
\end{equation}
where $v_{i}=dz_i/dt$ and $f$ denotes the constant pulling force. The kinetic friction is expressed as $-\nu_0-\nu_1v_i+\eta(i,z_i)$, where $\nu_0$ is a constant part of the kinetic friction and $-\eta(i,z_i)$ expresses a quenched randomness of the kinetic friction. In other words, $\nu_0-\eta(i,z_i)$ denotes the kinetic friction including quenched randomness owing to some randomness of the earth's crust. The term $-\nu_1v_i$ denotes the velocity-strengthening friction. The standard Coulomb friction, which does not depend on the velocity, corresponds to $\nu_1=0$.   If the inertial term represented by the left-hand side is neglected, Eq.~(8) becomes
\[
\nu_1\frac{dz_i}{dt}=k(z_{i+1}-2z_i+z_{i-1})+f-\nu_0+\eta(i,z_i).
\]
This equation has a form similar to Eq.~(1) when $\lambda=0$, that is, the qEW equation. 

We have performed a numerical simulation of Eq.~(8) by changing the external force $f$ at $k=0.5,\nu_1=1$, and $\nu_0=1$. The quenched randomness $\eta(i,z_i)$ is set to be a uniform random number between 0 and 0.5. The system size is $N=10000$. 
In other words, the kinetic friction takes a random number between 0.5 and 1. 
Equation (8) is applied to a moving block. If the block stops, a static friction force is applied to the block, and the block maintains the stationary state until the force $k(z_{i+1}-2z_i+z_{i-1})+f$ is beyond the maximum static friction. We assumed that the maximum static friction is $1$. 
Figure 2 shows the average velocity $\langle v\rangle=(1/N)\sum_{j=1}^N\langle dz_j/dt\rangle$ as a function of $f$. For $f<0.8125$, the average velocity is zero and all blocks stop finally. The average velocity increases continuously from 0 for $f>0.815$ owing to the velocity-strengthening coefficient $\nu_1>0$. When the viscous term is absent or $\nu_1=0$, the average velocity is zero for $f<0.749$, and the average velocity increases indefinitely for $f>0.749$. 

\begin{figure}[tbp]
\begin{center}
\includegraphics[height=3.5cm]{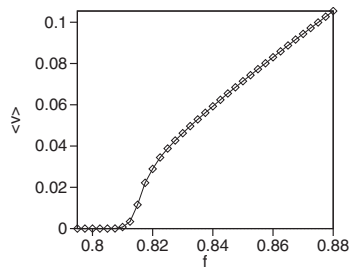}
\end{center}
\caption{Time averaged velocity $\langle v\rangle$ as a function of $f$  at $k=0.5$, $\nu_1=1$, and $\nu_0=1$ in Eq.~(8). }
\label{f2}
\end{figure}
\begin{figure}[tbp]
\begin{center}
\includegraphics[height=3.5cm]{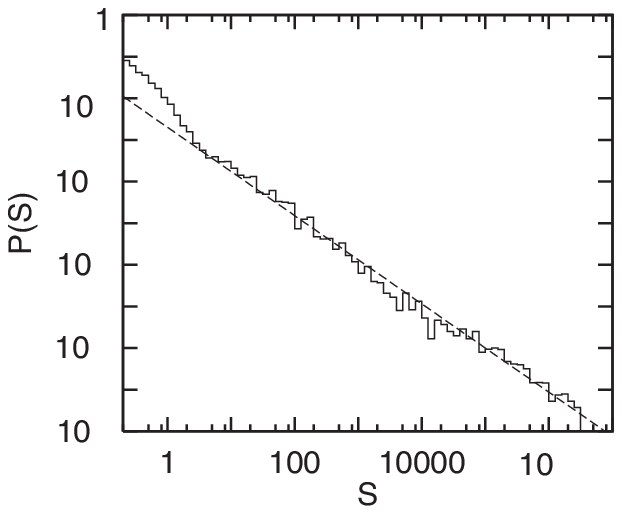}
\end{center}
\caption{Probability distribution $P(S)$ for $k=0.5$, $v_0=0.005$, $d=0.00005$, and $\nu_1=1$ in Eq.~(9). The dashed line denotes a power law of $\tau=1.06$. }
\label{f2}
\end{figure}
\begin{figure}[tbp]
\begin{center}
\includegraphics[height=3.5cm]{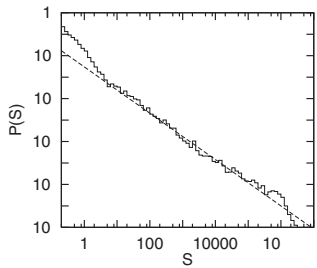}
\end{center}
\caption{Probability distribution $P(S)$ for $k=0.5$, $v_0=0.005$, $d=0.00005$, and $\nu_1=1$ in Eq.~(10). The dashed line denotes a power law of $\tau=1.08$.}
\label{f3}
\end{figure}
A slowly driven model can be expressed as
\begin{equation}
\frac{d^2z_i}{dt^2}=k(z_{i+1}-2z_i+z_{i-1})+v_0t-d\langle z\rangle-\nu_0-\nu_1v_{i}+\eta(i,z_i),
\end{equation}
where $\langle z\rangle=(1/N)\sum_{i=1}^Nz_i$ is the average value of the displacement $z_i$. The slowly driven model is appropriate in the block-spring model for earthquakes because plate motion is  very slow and has a steady velocity. 
 The parameters are set at $k=0.5,v_0=0.005$, $d=0.00005,\nu_0=1$, and $\nu_1=1$. The quenched randomness $\eta(i,z_i)$ takes a random number between 0 and 0.5. If the randomness is absent, all blocks move smoothly with a constant velocity as $z_i(t)=(v_0t-\nu_0-\nu_1v_0/d)/d$. In the numerical simulation, we have used the stepwise time evolution also for this block-spring model. That is, the numerical simulation of Eq.~(9) is performed for a fixed value of $v_0t-d\langle z\rangle$ until $dz_i/dt$ becomes 0 for every $i$. After the relaxation process, $v_0t$ is increased to $v_0(t+\Delta t)$. A sufficiently slow pulling  condition is realized in this type of numerical simulation. 
The effective pulling force $f=v_0t-d\langle z\rangle$ in Eq.~(9) fluctuates around a value of $f\sim 0.82$, which is close to the critical value of 0.815 in Eq.~(8).  
That is, a nearly critical state is self-organized in this model. 
Figure 3 shows the probability distribution of the slip size. The dashed line denotes a power law of $\tau=1.06$. The exponent is close to the exponent for the discretized qEW equation Eq.~(8). 

A slowly driven model without averaging is expressed as
\begin{equation}
\frac{d^2z_i}{dt^2}=k(z_{i+1}-2z_i+z_{i-1})+v_0t-dz_i-\nu_0-\nu_1v_{i}+\eta(i,z_i).
\end{equation}
The parameters are set at $k=0.5$, $v_0=0.005$, $d=0.00005$, $\nu_0=1$, and $\nu_1=1$. The quenched randomness $\eta(i,z_i)$ takes a random number between 0 and 0.5.  Figure 4 shows the probability distribution of the slip size. The dashed line denotes a power law of $\tau=1.08$. The exponent is also close to the exponent for the qEW equation. This implies that the averaging is not so important for the slip-size distributions.   

\begin{figure}[tbp]
\begin{center}
\includegraphics[height=3.5cm]{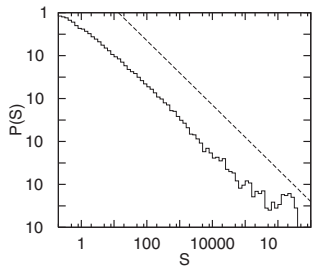}
\end{center}
\caption{Slip size distribution for $k=0.5$, $v_0=0.005$, $d=0.00005$, and $\nu_1=0$ in Eq.~(10). The dashed line denotes a power law of $\tau=1.5$.}
\label{f5}
\end{figure}
We have also performed  numerical results of Eq.~(10) for $\nu_1=0$, $k=0.5$, $v_0=0.005$, and $d=0.00005$. The coefficient $\nu_1$ of the velocity-strengthening friction is set to be 0. 
The time evolution of $f=v_0t-d\langle z\rangle$ shows that $f$ fluctuates around $f_c\sim 0.745$, which is close to the critical value $f_c\sim 0.749$ of Eq.~(8) at $\nu_1=0$. That is, a nearly critical state is also self-organized in this model equation. 
 Figure 5 shows the probability distribution $P(S)$. 
The dashed line denotes a power law of $1/S^{\tau}$ with $\tau\simeq 1.5$. For small slips, $P(S)$ obeys a power law of $\tau\sim 1.5$. However, a small hump structure appears around $S\sim 2\times 10^{6}$, which corresponds to the rather regular large-scale slips.  We do not yet understand the reason for the deviation from the pure power law, but we conjecture that it might originate from the fact that the transition at $f_c\sim 0.749$ in Eq.~(8) at $\nu_1=0$ is not a continuous transition such as the second-order phase transition, but the velocity increases indefinitely for $f>f_c$ in the time evolution of Eq.~(8).     
  
\begin{figure}[tbp]
\begin{center}
\includegraphics[height=3.5cm]{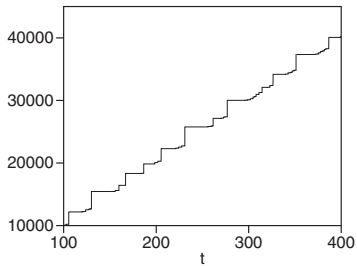}
\end{center}
\caption{Time evolution of the average $\langle z\rangle=(1/N)\sum_{j=1}^Nz_j$ in Eq.~(11) at $\gamma=0.001$. }
\label{f6}
\end{figure}
\begin{figure}[tbp]
\begin{center}
\includegraphics[height=3.5cm]{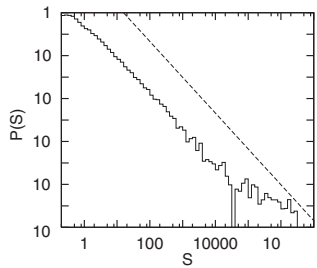}
\end{center}
\caption{Slip-size distributions for Eq.~(11) at $\gamma=0.001$. The dashed line denotes a power law of $\tau=1.67$.}
\label{f7}
\end{figure}
If the velocity-weakening friction proposed by Carlson and Langer is assumed, the model equation is written as
\begin{equation}
\frac{d^2z_i}{dt^2}=k(z_{i+1}-2z_i+z_{i-1})+v_0t-dz_i-\frac{\nu_0-\eta(i,z_i)}{1+\gamma v_{zi}}.
\end{equation}
Figures 6 shows the time evolution of $\langle z\rangle$ at $0.001$. Very large stick-slip motion becomes clearer as $\gamma$ is increased.  Figure 7 shows the slip-size distributions at $\gamma=0.001$. The distribution can be approximated as a power law of $\tau\sim  1.67$ for small slips, but large slips occur more frequently. The exponent $\tau=1.67$ is close to the exponent of the Gutenberg-Richter law. 
\section{Summary}
We have investigated block-spring models with quenched randomness. If the velocity-strengthening friction is assumed, a kind of SOC is realized when the pulling velocity is sufficiently small. 
The slip-size distribution obeys a power law of exponent $\tau\sim 1.08$ in the one-dimensional model, which is close to the exponent for the quenched Edwards-Wilkinson equations. This suggests that our velocity-strengthening block-spring model with quenched randomness belongs to the same universality class as the qEW equation. 

We have investigated the slip-size distributions in more general cases such as  Coulomb friction and velocity-weakening friction. The occurrence of SOC is questionable for these general cases because some deviation from the power law is observed for very large slips; however, power laws are observed for small size slips. The exponents are evaluated to be $\tau=1.5$ for the Coulomb friction and $\tau=1.67$ for the velocity-weakening friction. 
This power-law behavior for small size slips was not observed in the homogeneous Carlson-Langer model for $\gamma<<2$. The quenched randomness plays an important role even in the velocity weakening friction.


\begin{thebibliography}{99}
\bibitem{rf:1} B.~Gutenberg and C.~F.~Richter, {\it Seismicity of the Earth and Related Phenomena} (Princeton University Press, Princeton, NJ, 1954).
\bibitem{rf:2} C.~H.~Scholz, {\it The Mechanics of Earthquakes and Faulting} (Cambridge University Press, Cambridge, 2002).
\bibitem{rf:3} R.~Burridge and L.~Knopoff, Bull. Seismol. Soc. Am. {\bf 57}, 341 (1967).
\bibitem{rf:4} J.~M.~Carlson and J.~S.~Langer, Phys. Rev. Lett. {\bf 62}. 2632 (1989).
\bibitem{rf:5} T.~Cao and K.~Aki, Pure. Appl. Geophys. {\bf 124}, 487 (1986).
\bibitem{rf:6} M.~de Sousa Vieira, G.~L.~Vasconcelos, and S.~R.~Nagel,  Phys. Rev. E {\bf 47}, R2221 (1993).
\bibitem{rf:7} T.~Mori and H.~Kawamura, J. Geophys. Res. {\bf 113}, B06301 (2008).
\bibitem{rf:8} H.~Sakaguchi and S.~Morita, J. Phys. Soc. Jpn. {\bf 82}, 114006 (2013).
\bibitem{rf:9} D.~P.~Vallette and J.~P.~Gollub, Phys. Rev. E {\bf 47}, 820 (1993).
\bibitem{rf:10} T.~Yamaguchi, M.~Morishita, M.~Doi, T.~Hori, H.~Sakaguchi, and J.~P.~Ampuero, J.~Geophys. Res. {\bf 116}, B12306 (2011).
\bibitem{rf:11} P.~Bak, P.~Tang, and K.~Wiesenfeld, Phys. Rev. Lett. {\bf 59}, 381 (1989).
\bibitem{rf:12} V.~Frette, K.~Christensen, A.~Mathe-S$\o$renssen, J.~Feder, T.~J$\o$ssang, and P.~Meakin, Nature {\bf 379}, 49 (1996).
\bibitem{rf:13} M.~A.~Rubio, C.~A.~Edwards, A.~Dougherty, and J.~P.~Gollub, Phys. Rev. Lett. {\bf 63}, 1685 (1989).
\bibitem{rf:14} T.~Vicsek, M.~Cerzo, and V.~K.~Horv\'ath, Physica A {\bf 167}, 315 (1990).
\bibitem{rf:15} Z.~Csah\'ok, K.~Honda, and T.~Vicsek. J. Phys. A {\bf 26}, L171 (1993).
\bibitem{rf:16} H.~Leschhorn, Phys. Rev. E {\bf 54}, 1313 (1996).
\bibitem{rf:17} M.~Kardar, G.~Parisi, and Y.-C.~Zhang, Phys. Rev. Lett. {\bf 56}, 889 (1986).
\bibitem{rf:18} S.~F.~Edwards and D.~R.~Wilkinson, Proc. R. Soc. London A {\bf 381}, 1 (1982).
\bibitem{rf:19} M.~Paczuski and S.~Boettcher, Phys. Rev. Lett. {\bf 77}, 111 (1996).
\bibitem{rf:20}  S.~V.~Buldyrev, A.-L.~Barab\'asi, F.~Caserta, S.~Havlin, H.~E.~Stanley, and T.~Vicsek, Phys. Rev. A {\bf 45}, R8313 (1992).
\bibitem{rf:21} H.~Sakaguchi, Phys. Rev. E {\bf 82}, 032101 (2010).
\bibitem{rf:22} F.~Lacobe, S.~Zapperi, and H.~J.~Herrmann, Phys. Rev. B {\bf 63}, 104104 (2001).
\bibitem{rf:23} Y.-J.~Chen, S.~Papanikolaou, J.~P.~Sethna, S.~Zappero, and G.~Durin, Phys. Rev. E {\bf 84}, 061103 (2011).
\end{thebibliography}
\end{document}